\documentclass{jetpl}
\twocolumn

\RequirePackage{amsmath}
\RequirePackage{amssymb}
\RequirePackage{lineno}
\RequirePackage[colorlinks,citecolor=blue,urlcolor=blue,linkcolor=blue]{hyperref}

\lat

\title{Neutral bremsstrahlung electroluminescence in noble liquids revisited}

\rtitle{Neutral bremsstrahlung electroluminescence\ldots}

\sodtitle{Neutral bremsstrahlung electroluminescence in noble liquids revisited}

\author{A.\,F.\,Buzulutskov$^{+*}$,
E.\,A.\,Frolov$^{+*}$\/\thanks{e-mail: geffdroid@gmail.com}
}

\rauthor{A.\,F.\,Buzulutskov, E.\,A.\,Frolov}

\sodauthor{Buzulutskov, Frolov}

\address{$^+$Budker Institute of Nuclear Physics SB RAS, Lavrentiev avenue 11, 630090 Novosibirsk, Russia\\~\\
$^*$Novosibirsk State University, Pirogova street 2, 630090 Novosibirsk, Russia}

\dates{---}{*}

\abstract{Recent discovery of neutral bremsstrahlung (NBrS) mechanism of electroluminescence (EL) in noble gases in two-phase detectors for dark matter searches has led to a prediction that NBrS EL should be present in noble liquids as well. A rigorous theory of NBrS EL in noble liquids was developed accordingly in the framework of Cohen-Leckner and Atrazhev formalism. It has been recently followed by the first experimental observation of NBrS EL in liquid argon, which however deviates significantly from the previous theory. Given these results, we revise previous theoretical calculations of EL NBrS in noble liquids to be consistent with experiment. In particular, NBrS EL yield and spectra were calculated in this work for argon, krypton and xenon with momentum-transfer cross section of electron scattering (instead of energy-transfer one) being used for calculation of NBrS cross section. The results for light noble liquids, helium and neon, are also reexamined.}


\begin{document}

\maketitle
\section{Introduction}\label{intro}

Electroluminescence (EL) is a physical phenomenon in which medium produces light in response to applied electric field or electric current. EL in noble gases is a crucial effect used in two-phase (liquid-gas) detectors for low-energy and low-background experiments such as dark matter searches and coherent neutrino-nucleus scattering~\cite{Akimov21,Aprile18,Agnes18b,LUX16}. In these detectors, both prompt scintillation (S1) signal and delayed primary ionization (S2) signal are measured, the latter being recorded using the EL effect in the gas phase.

Recently, a new mechanism of EL, namely that of neutral bremsstrahlung (NBrS) emission in the UV, visible and near infrared (NIR) range~\cite{Buzulutskov18,Borisova21,Borisova22,Henriques22,Aoyama22,Milstein22,Milstein23}, was discovered in gaseous Ar~\cite{Buzulutskov18} and Xe~\cite{Henriques22}. This EL is due to elastic scattering of electrons on neutral atoms:
\begin{equation}
\label{NBrS_react}
e^- + \mathrm{Ar} \rightarrow e^- + \mathrm{Ar} + h\nu \; . 
\end{equation}
While NBrS EL yield is small compared to other EL mechanisms~\cite{Akimov21,Borisova21,Buzulutskov20}, NBrS EL has no threshold in the electric field and has an advantage of wide emission spectrum, from UV to NIR, suitable for direct optical readout by conventional photodetectors such as photomultiplier tubes (PMTs) and silicon photomultipliers (SiPMs).

Following the discovery of NBrS EL in noble gases, its presence in liquids was also predicted theoretically~\cite{Borisova22}, and later observed for the first time in Ar~\cite{Buzulutskov23b}. Besides obvious interest in NBrS as a new physical phenomenon, interest to it both in gases and liquids is also justified by new and more reliable readout schemes of two-phase and single-phase noble liquid detectors which may be developed based on this effect.

\begin{figure}[!t]
	\centering
	\includegraphics[width=1.0\linewidth]{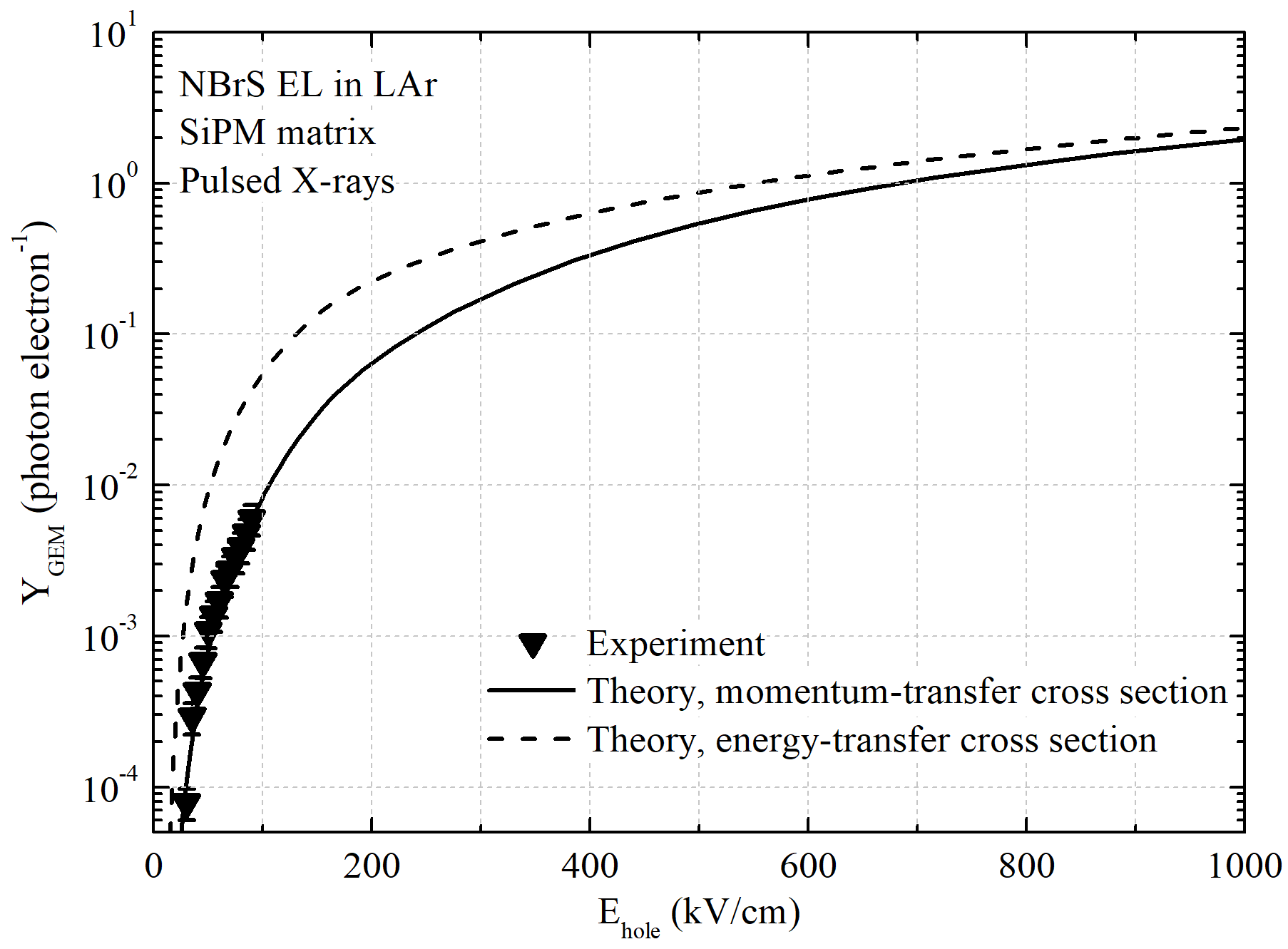}
	\caption{\textbf{Fig.~\ref*{fig:GEM_yield}} Absolute light yield of NBrS EL in liquid Ar produced in Gas Electron Multiplier (GEM) structure~\cite{Sauli16} and expressed in photons (at $\lambda$$\le$1000 nm) per drifting electron as a function of the electric field in GEM hole center. Theoretical predictions for both when momentum-transfer and when energy-transfer cross section is used to calculate NBrS cross section (see Eq.~\ref{NBrS-XS}) are shown by solid and dashed line respectively. The experimental data and theoretical predictions were obtained in~\cite{Buzulutskov23b}.}
	\vspace{-10pt}
	\label{fig:GEM_yield}
\end{figure}

This work is reexamination of theoretical work~\cite{Borisova22}, where NBrS yields and spectra were calculated for all noble liquids from He to Xe, in the light of latest experimental results on EL in liquid Ar~\cite{Buzulutskov23b}. In particular, there was an ambiguity as to whether energy-transfer (total elastic) or momentum-transfer (transport) cross section of electron scattering on effective potential should be used to calculate NBrS cross section (see section~\ref{theory}~). The energy-transfer one was used in~\cite{Borisova22}, whereas the experimental results~\cite{Buzulutskov23b} unambiguously show that using momentum-transfer cross section is the correct choice (see Fig.~\ref{fig:GEM_yield}). Thus, in this work we recalculate NBrS EL for heavy gases (Ar, Kr and Xe) according to Cohen-Lekner~\cite{Cohen67} and Atrazhev~\cite{Atrazhev85} approach in the same manner as was done in~\cite{Borisova22}. Results on light elements (He and Ne) which were calculated in~\cite{Borisova22} using approximation where liquid was taken as gas with appropriate atomic density (``compressed gas'' approximation) are also discussed. 

\section{Theory}\label{theory}

The calculations of NBrS EL performed in~\cite{Borisova22} are based on Cohen-Lekner~\cite{Cohen67} and Atrazhev~\cite{Atrazhev85} formalism. In this approach the electron transport through the liquid is considered as a sequence of single scatterings described by scattering cross sections with a distinction between the energy transfer scattering, which changes only the electron energy, and that of momentum transfer, which changes only the direction of the electron velocity. Both processes are assigned separate cross sections~\cite{Cohen67,Atrazhev85,Stewart10}. Those given in~\cite{Atrazhev85} are used in this work, since then the theory describes well the experimental data on the electron drift velocities~\cite{Borisova22,Atrazhev85}. For calculating NBrS EL, only distribution of drifting electrons by their energy $f(\varepsilon)$ and their drift velocity $v_d$ is required.

Besides electron transport parameters, formula of cross section of NBrS photon emission due to electron scattering is also required. Its compact form~\cite{Buzulutskov18,Henriques22,Firsov60,Kasyanov65,Dalgarno66,Biberman67,Park00} is the following:

\begin{eqnarray}
\frac{d\sigma}{d\nu} = \frac{8}{3} \frac{r_e}{c} \frac{1}{h\nu}&& \left(\frac{\varepsilon-h\nu}{\varepsilon}\right)^{1/2} \nonumber\\ &&\times [(\varepsilon-h\nu)\sigma_{el}(\varepsilon) + \varepsilon\sigma_{el}(\varepsilon - h\nu)]\;, \label{NBrS-XS}
\end{eqnarray}
where $h\nu$ is the photon energy, $r_e = e^2/mc^2$ is the classical electron radius, $c$ is the speed of light, $\varepsilon$ is the energy of incident electron and $\sigma_{el}(\varepsilon)$ is the cross section of its elastic scattering on effective potential. It was proposed in~\cite{Borisova22} that within the  Cohen-Lekner~\cite{Cohen67} and Atrazhev~\cite{Atrazhev85} approach this formula, initially derived for gas, can also be applied for liquid.

The absolute EL yield, $Y_{EL}$, is defined as the number of photons produced per unit drift path and per drifting electron. To compare results at different medium densities and temperatures, reduced EL yield, $Y_{EL}/N$, is used instead, where $N$ is liquid atomic density. It is as a function of reduced electric field $E/N$ expressed in Td units (1~Td$=10^{-17}$~V~cm$^2$). Given the electron energy distribution function $f(\varepsilon)$ normalized to unity, it can be described by the following equation~\cite{Buzulutskov18}:
\begin{equation}
\label{NBrS-Y}
\frac{Y_{\mathrm{EL}}}{N} = \int_{\lambda_{1}}^{\lambda_{2}}\!\!\! \int_{h\nu}^{\infty}\!\!\frac{\sqrt{2\varepsilon/m}}{v_d} \frac{d\sigma}{d\nu}\frac{d\nu}{d\lambda}\, f(\varepsilon)\, d\varepsilon\, d\lambda \;, 
\end{equation}
where $\lambda_1$--$\lambda_2$ is the wavelength region of interest. In this work as well as in our previous ones it is limited to the wavelength region of 0-1000~nm. The wavelength spectra of the NBrS EL can be obtained from this equation by taking its derivative with respect to~$\lambda$. Calculations were made for liquids at boiling temperature at 1.0~atm pressure, atomic densities being taken from~\cite{Fastovsky72}.

\begin{figure}[!t]
\includegraphics[width=1.0\linewidth]{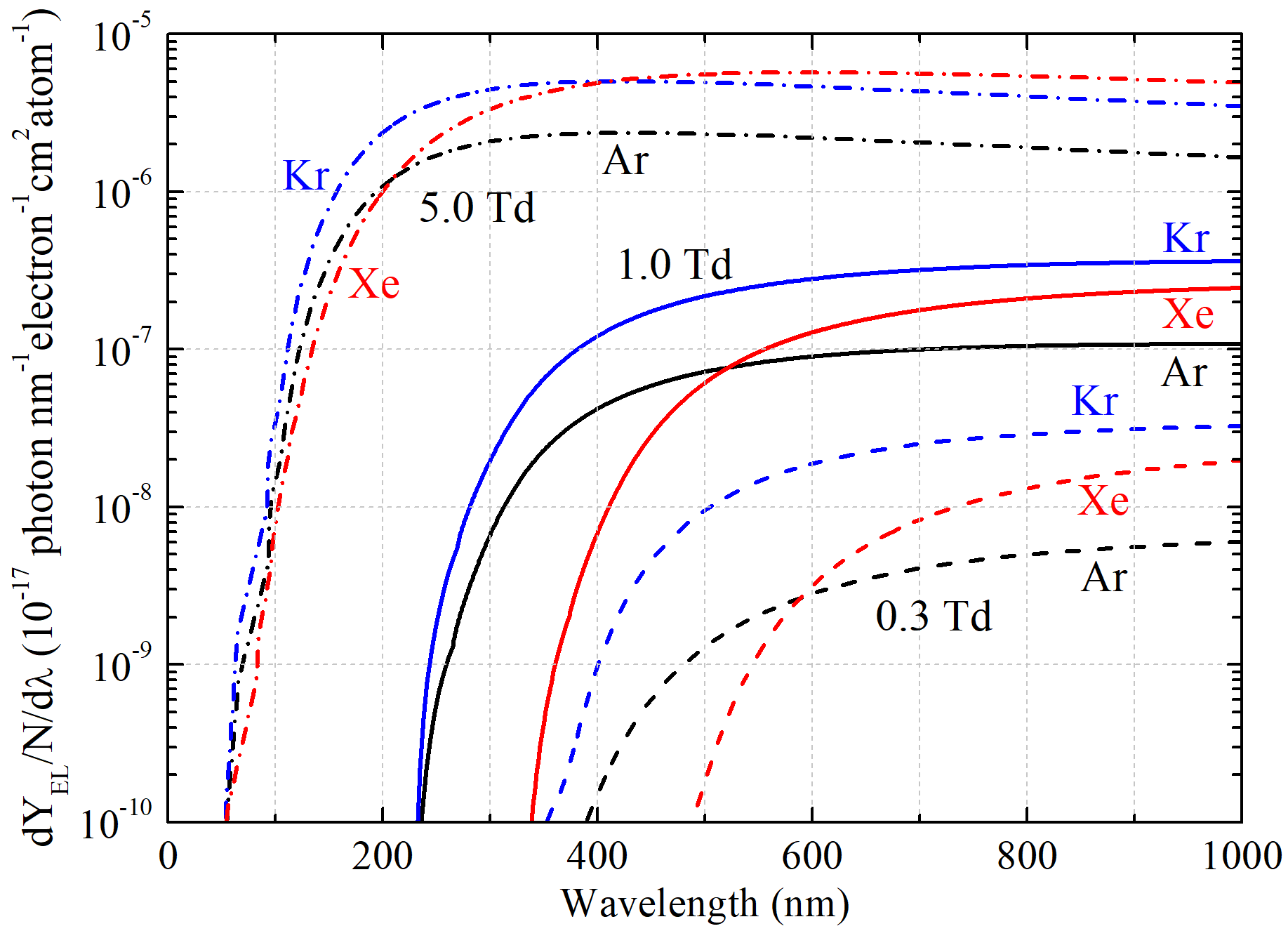}
\caption{\textbf{Fig.~\ref*{fig:spectra}} Theoretical spectra of the reduced EL yield for NBrS EL in liquid Ar, Kr and Xe at different reduced electric fields. The results are obtained with the momentum-transfer cross section (taken from~\cite{Atrazhev85}) being used in Eq.~(\ref{NBrS-XS}).}
\label{fig:spectra}
\end{figure}

\begin{figure}[!t]
\includegraphics[width=1.0\linewidth]{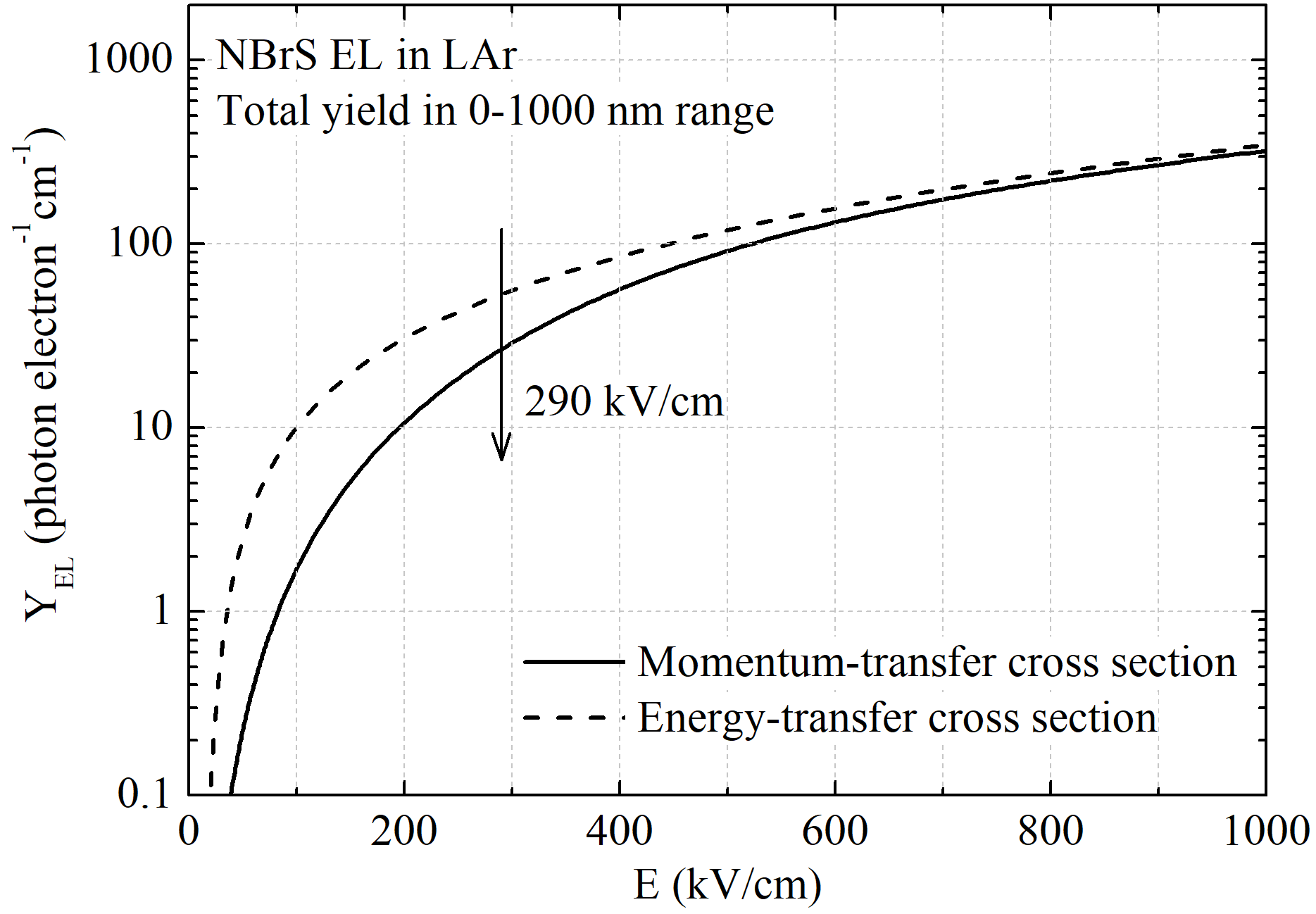}
\caption{\textbf{Fig.~\ref*{fig:yields_Ar}} Theoretic absolute EL yield for NBrS EL in liquid Ar at 0-1000~nm as a function of electric field. The results are obtained for both the momentum-transfer and energy-transfer cross section being used in Eq.~(\ref{NBrS-XS}). Both cross sections are taken from~\cite{Atrazhev85}. The arrow indicates electric field at which twofold difference between using the momentum-transfer cross section and that of energy-transfer (the latter being equivalent to compressed gas approximation) is reached. }
\label{fig:yields_Ar}
\end{figure}
\begin{figure}[!t]
\includegraphics[width=1.0\linewidth]{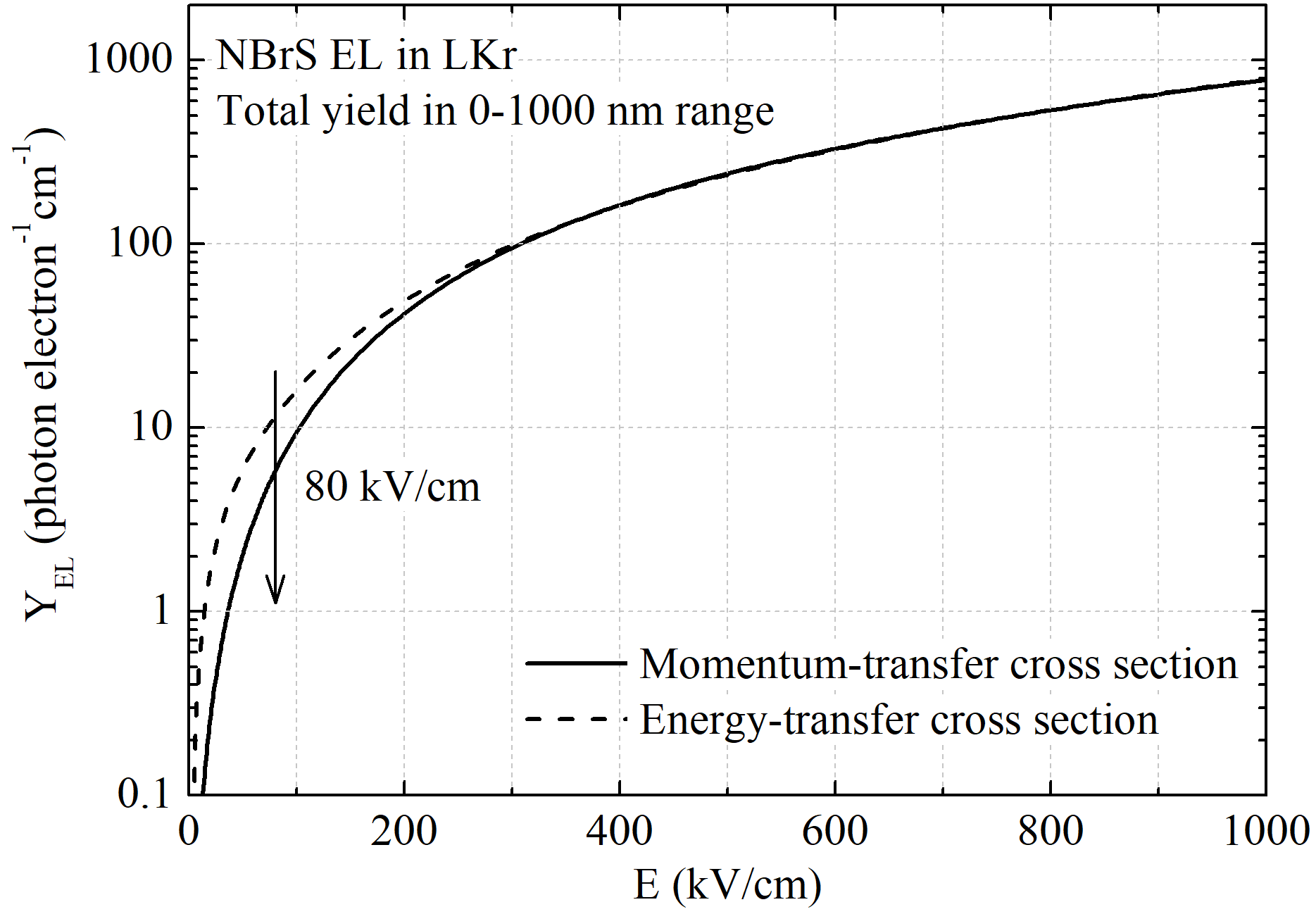}
\caption{\textbf{Fig.~\ref*{fig:yields_Kr}} Same as Fig.~\ref{fig:yields_Ar} done for Kr.}
\label{fig:yields_Kr}
\end{figure}
\begin{figure}[!t]
\includegraphics[width=1.0\linewidth]{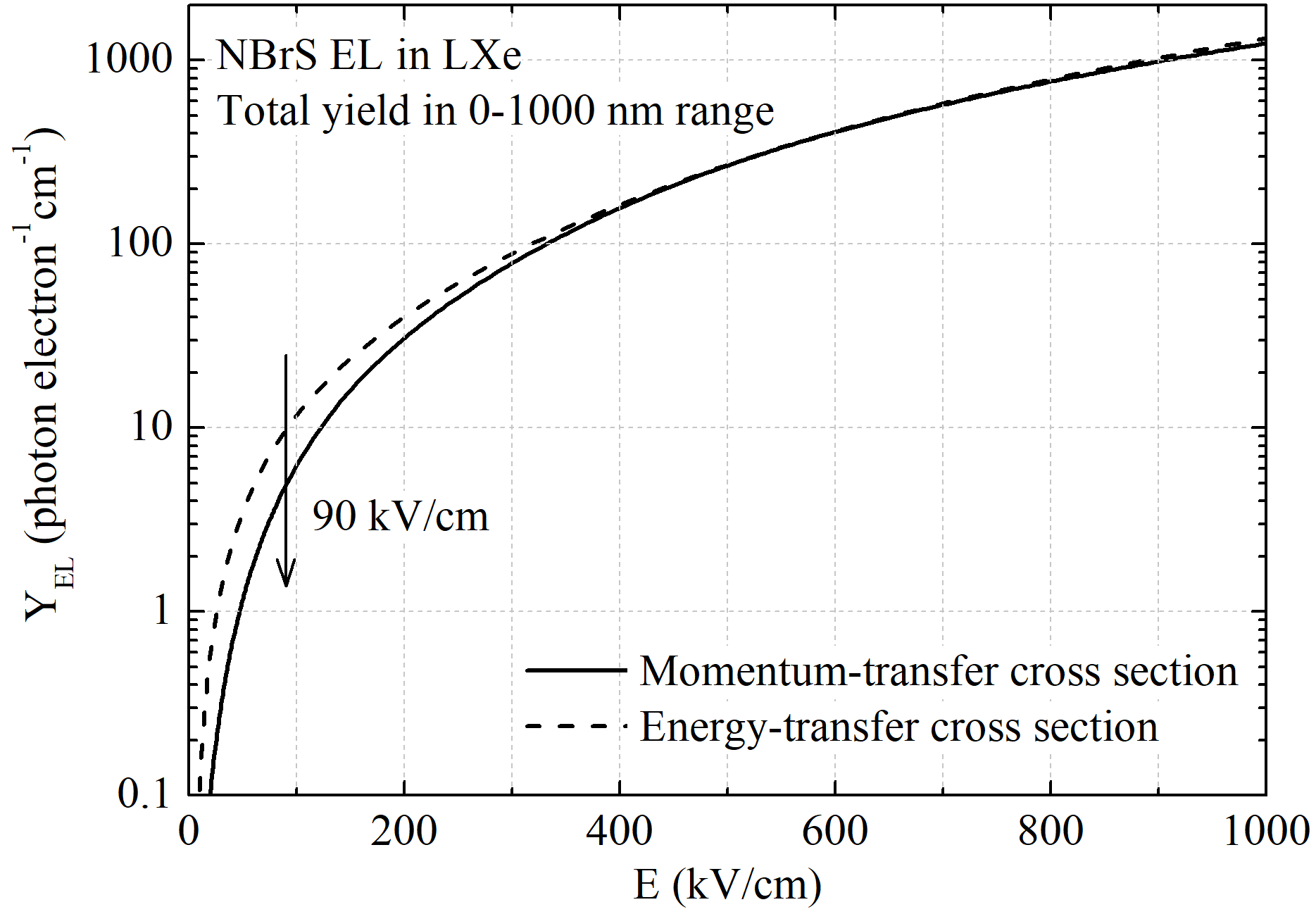}
\caption{\textbf{Fig.~\ref*{fig:yields_Xe}} Same as Fig.~\ref{fig:yields_Ar} done for Xe.}
\label{fig:yields_Xe}
\end{figure}

It should be remarked that until recently there was an ambiguity about what cross section of elastic scattering, $\sigma_{el}(\varepsilon)$, should appear in Eq.~(\ref{NBrS-XS}): some theoretical derivations show that it should be energy-transfer (total elastic) cross section~\cite{Firsov60,Kasyanov65,Dalgarno66,Biberman67,Kasyanov78}, while others show that it should be momentum-transfer (transport) one~\cite{Milstein22,Dalgarno66,Biberman67,Park00}. And while in gases these two approaches do not result in significantly different NBrS yields~\cite{Buzulutskov18,Henriques22,Amedo22}, in~\cite{Buzulutskov23b} and this work it is shown that in liquids the results can differ by more than an order of magnitude, depending on the electric field. Experimental measurements of absolute EL yields in liquid Ar (see Fig.~\ref{fig:GEM_yield} and~\cite{Buzulutskov23b}) put this question to rest and showed that using momentum-transfer cross section is the correct choice. It should be noted that the measurements of NBrS EL in gaseous Xe~\cite{Henriques22} also favored the use of momentum-transfer cross section although is was difficult to draw a convincing conclusion given experimental uncertainties there.

Therefore, recalculation of NBrS EL results from~\cite{Borisova22} for Ar, Kr and Xe was done using momentum-transfer cross section and the results are shown in Figs.~\ref{fig:spectra},~\ref{fig:yields_Ar},~\ref{fig:yields_Kr} and~\ref{fig:yields_Xe}. Fig.~\ref{fig:spectra} shows the wavelength spectra of the NBrS EL at different electric fields. One can see that they are rather flat and do not differ much in shape from those of noble gases~\cite{Borisova21}. Their shape also does not depend strongly on whether momentum-transfer or energy-transfer cross section is used in Eq.~(\ref{NBrS-XS}) (compare Fig.~\ref{fig:spectra} here to Fig.~3 in~\cite{Borisova22}).

Figs.~\ref{fig:yields_Ar},~\ref{fig:yields_Kr} and~\ref{fig:yields_Xe} show absolute EL yields as a function of the electric field and compare their values obtained with momentum-transfer and energy-transfer cross sections. As can be seen, two cross sections result in difference of yields of about an order of magnitude at low electric fields, below 100 kV/cm, most relevant to experiments with GEMs and Thick GEMs. As the field increases, however, this difference decreases and almost disappears at sufficiently high electric fields, exceeding several hundreds kV/cm. Such strong electric fields can be obtained in noble liquids with thin wires~\cite{Aprile14b} or needles~\cite{Schussler00}. This means that for higher electric fields the results of~\cite{Borisova22} on NBrS EL in heavy noble gases remain valid.  

It must be remarked that for light noble liquids, He and Ne, the theory described above cannot be applied because appropriate cross sections for electron transport in liquid are not available in the literature. In~\cite{Borisova22} this problem was avoided by using ``compressed gas'' approximation. The basis of this approximation is that in gases, electron distribution functions $f(\varepsilon)$ and reduced EL yields $Y_{EL}/N$ are actually functions of not electric field $E$ but of reduced electric field $E/N$. This allows to scale calculations conducted at one density to any other, strictly speaking until effective potential of electron scattering sufficiently changes. In~\cite{Borisova22} it was shown that, surprisingly, extending this scaling to liquid densities resulted in the same, within factor of 2, NBrS yields for Ar, Kr and Xe, as those calculated in Cohen-Lekner and Atrazhev formalism using energy-transfer cross section.

In~\cite{Borisova22}, this justified the use of compressed gas approximation for light elements, He and Ne, as well. However, given that momentum-transfer cross section must be used instead for the exact solution, compressed gas approximation becomes invalid at low fields (see Figs.~\ref{fig:yields_Ar},~\ref{fig:yields_Kr} and~\ref{fig:yields_Xe}). In particular, the electric fields where difference between using momentum-transfer cross section and energy-transfer one becomes less than twofold are above 290~kV/cm (1.4~Td), 80~kV/cm (0.5~Td) and 90~kV/cm (0.7~Td) for Ar, Kr and Xe respectively. Corresponding reduced electric fields are given in parentheses. It may be then expected from extrapolation that for He and Ne the compressed gas approximation would work only for fields higher than 1.4~Td, or 265 and 500~kV/cm correspondingly.

\section{Conclusions}

Given new experimental data on EL in liquid Ar~\cite{Buzulutskov23b}, recalculation of NBrS EL in noble liquids, carried out for the first time in~\cite{Borisova22}, was conducted in this work. In particular, NBrS EL spectra and yields were obtained for liquid Ar, Kr and Xe in a wide range of electric fields using momentum-transfer cross section of electron elastic scattering on effective potential for calculating NBrS cross section instead of previously used energy-transfer one. It was found that NBrS yields can be significantly smaller, by an order of magnitude, than previously thought for low electric fields, below 100~kV/cm. On the other hand, this difference decreases with field and almost disappears at sufficiently high electric fields, exceeding 290, 80 and 90~kV/cm for Ar, Kr and Xe respectively.

Accordingly, the conclusion that NBrS EL may find applications for alternative readout concepts for single-phase noble liquid detectors remains valid. 

Finally, it was also shown that NBrS EL yields for He and Ne can be calculated using compressed gas approximation within uncertainty of a factor of 2 for electric fields exceeding 265 and 500~kV/cm respectively.

\section*{Acknowledgments}
This work was supported in part by Russian Science Foundation (project no. 20-12-00008, \url{https://rscf.ru/project/20-12-00008/}).

\bibliographystyle{jetpl}
\bibliography{Manuscript}

\newcommand{\JINST}{J. Instrum.}\newcommand{\PhysLett}{Phys.
  Lett.}\newcommand{\IEEETransElectronDevices}{IEEE Trans. Electron
  Devices}\newcommand{\IEEETransNuclSci}{IEEE Trans. Nucl.
  Sci.}\newcommand{\AstropartPhys}{Astropart. Phys.}\newcommand{\NIM}{Nucl.
  Instrum. Methods}\newcommand{\NIMA}{Nucl. Instrum. Meth.
  A}\newcommand{\NIMB}{Nucl. Instrum. Meth. B}\newcommand{\JApplPhys}{J. Appl.
  Phys.}\newcommand{\EPJWebConf}{EPJ Web Conf.}\newcommand{\EPJC}{Eur. Phys. J.
  C}\newcommand{\PhysRevLett}{Phys. Rev. Lett.}\newcommand{\PhysRevD}{Phys.
  Rev. D}\newcommand{\InstrumExpTech}{Instrum. Exp.
  Tech.}\newcommand{\InstrumMDPI}{Instruments}\newcommand{\JPAC}{J. Cosmol.
  Astropart. Phys.}\newcommand{\EPL}{Europhys. Lett.}
\begin{thebibliography}{10}

\bibitem{Akimov21}
D.~Y. Akimov, A.~I. Bolozdynya, A.~F. Buzulutskov, and V.~Chepel, Two-Phase
  Emission Detectors,  (World Scientific2021).

\bibitem{Aprile18}
E.~Aprile et~al., Phys. Rev. Lett. \textbf{121}, 111302 (2018).

\bibitem{Agnes18b}
P.~Agnes et~al., Phys. Rev. D \textbf{98}, 102006 (2018).

\bibitem{LUX16}
D.~S. Akerib et~al., Phys. Rev. Lett. \textbf{116}, 161301 (2016).

\bibitem{Buzulutskov18}
A.~Buzulutskov, E.~Shemyakina, A.~Bondar, A.~Dolgov, E.~Frolov, V.~Nosov,
  V.~Oleynikov, L.~Shekhtman, and A.~Sokolov, Astropart. Phys. \textbf{103}, 29
  (2018).

\bibitem{Borisova21}
E.~Borisova and A.~Buzulutskov, Eur. Phys. J. C \textbf{81}, 1128 (2021).

\bibitem{Borisova22}
E.~Borisova and A.~Buzulutskov, Europhys. Lett. \textbf{137}, 24002 (2022).

\bibitem{Henriques22}
C.~A.~O. Henriques et~al., Phys. Rev. X \textbf{12}, 021005 (2022).

\bibitem{Aoyama22}
K.~Aoyama, M.~Kimura, H.~Morohoshi, T.~Takeda, M.~Tanaka, and K.~Yorita, Nucl.
  Instrum. Meth. A \textbf{1025}, 166107 (2022).

\bibitem{Milstein22}
A.~Milstein, S.~Salnikov, and M.~Kozlov, Nucl. Instrum. Meth. B \textbf{530},
  48 (2022).

\bibitem{Milstein23}
A.~Milstein, S.~Salnikov, and M.~Kozlov, Nucl. Instrum. Meth. B \textbf{539}, 9
  (2023).

\bibitem{Buzulutskov20}
A.~Buzulutskov, Instruments \textbf{4}, 16 (2020).

\bibitem{Buzulutskov23b}
A.~Buzulutskov, E.~Frolov, E.~Borisova, V.~Nosov, V.~Oleynikov, and A.~Sokolov,
  First observation of neutral bremsstrahlung electroluminescence in liquid
  argon (2023), eprint arXiv: 2305.08084.

\bibitem{Sauli16}
F.~Sauli, Nucl. Instrum. Meth. A \textbf{805}, 2 (2016), special Issue in
  memory of Glenn F. Knoll.

\bibitem{Cohen67}
M.~H. Cohen and J.~Lekner, Phys. Rev. \textbf{158}, 305 (1967).

\bibitem{Atrazhev85}
V.~M. Atrazhev and E.~G. Dmitriev, J. Phys. C \textbf{18}, 1205 (1985).

\bibitem{Stewart10}
D.~Y. Stewart et~al., J. Instrum. \textbf{5}, P10005 (2010).

\bibitem{Firsov60}
O.~B. Firsov and M.~I. Chibisov, Sov. Phys. JETP \textbf{12}, 1235 (1960).

\bibitem{Kasyanov65}
V.~A. Kas’yanov and A.~N. Starostin, Sov. Phys. JETP \textbf{21}, 193 (1965).

\bibitem{Dalgarno66}
A.~Dalgarno and N.~F. Lane, Astrophys. J. \textbf{145}, 623 (1966).

\bibitem{Biberman67}
L.~M. Biberman and G.~E. Norman, Sov. Phys. Usp. \textbf{10}, 52 (1967).

\bibitem{Park00}
J.~Park, I.~Henins, H.~W. Herrmann, and G.~S. Selwyn, Phys. Plasmas \textbf{7},
  3141 (2000).

\bibitem{Fastovsky72}
V.~G. Fastovsky, A.~E. Rovinsky, and Y.~V. Petrovsky, Inert Gases,  (Moscow
  Atomizdat (in Russian)1972), second ed.

\bibitem{Kasyanov78}
V.~A. Kas'yanov and A.~N. Starostin, Sov. J. Plasma Phys. \textbf{4}, 67
  (1978).

\bibitem{Amedo22}
P.~Amedo, D.~Gonzalez-Díaz, and B.~Jones, J. Instrum. \textbf{17}, C02017
  (2022).

\bibitem{Aprile14b}
E.~Aprile, H.~Contreras, L.~W. Goetzke, A.~J.~M. Fernandez, M.~Messina,
  J.~Naganoma, G.~Plante, A.~Rizzo, P.~Shagin, and R.~Wall, J. Instrum.
  \textbf{9}, P11012 (2014).

\bibitem{Schussler00}
A.~S. Schussler, J.~Burghorn, P.~Wyder, B.~I. Lembrikov, and R.~Baptist, Appl.
  Phys. Lett. \textbf{77}, 2786 (2000).

\end{thebibliography}

\end{document}